\documentclass[twocolumn,showpacs,preprintnumbers,amsmath,amssymb]{revtex4}
\usepackage{dcolumn}% Align table columns on decimal point
\usepackage{bm}% bold math
\newcommand{\be}{\begin{equation}}  
\newcommand{\ee}{\end{equation}}  
\newcommand{\bear}{\begin{eqnarray}}  
\newcommand{\eear}{\end{eqnarray}}  
\newcommand{\ba}{\begin{array}}  
\newcommand{\ea}{\end{array}}

\newskip\humongous \humongous=0pt plus 1000pt minus 1000pt

\newif\ifdtup

\def\oldreffmt#1{\rlap{[#1]} \hbox to 2\parindent{}}

\def\figfmt#1{\rlap{Figure {#1}} \hbox to 1in{}}  
  
\def\ie{\hbox{\it i.e.}{}}        
\def\eg{\hbox{\it e.g.}{}}

\def \Tr{\mathop{\rm Tr}}

\def\slash#1{#1\!\!\!/\!\,\,}  
\def\beq{\begin{equation}}  
\def\eeq{\end{equation}}  
\def\bea{\begin{eqnarray}}  
\def\eea{\end{eqnarray}}

\def\bq{\begin{quote}}  
\def\eq{\end{quote}}

\relax  
\newdimen\tdim  
\tdim=\unitlength  
\def\bar{\overline}

\begin{document}  

\preprint{FERMILAB-Pub-08-027-T \qquad ANL-HEP-PR-08-8} 
\title{
%{\Huge Preliminary v1.10}\\ \vspace{0.2in}
Chern-Simons and WZW Anomaly Cancelations Across Dimensions}

\author{Christopher T. Hill}
{\email{hill@fnal.gov}}
\affiliation{
 {{Fermi National Accelerator Laboratory}}\\
{{\it P.O. Box 500, Batavia, Illinois 60510, USA}}
}%

\author{Cosmas K. Zachos}
{\email{zachos@anl.gov}}
\affiliation{ {{High Energy Physics Division,
Argonne National Laboratory}}\\
{{\it  Argonne, IL 60439-4815, USA}}
}%
\date{February 12, 2008}% It is always \today, today,

\begin{abstract} 
The WZW functional in $D=4$ can be derived directly from the Chern-Simons
functional of a compactified $D=5$ gauge theory and the boundary fermions
it supplants. A simple pedagogical model
based on $U(1)$ gauge groups illustrates how this works. A bulk-boundary
system with the fermions eliminated manifestly evinces anomaly 
cancelations between CS and WZW terms.
 \end{abstract}

\pacs{11.15.-q,12.15.-y,12.38.Qk,12.39.Fe,
13.15.+g,13.40.-f,14.70.Hp,14.80.Mz,95.85.Ry,97.60.Jd}
\maketitle

\section{Introduction}

In this paper we illustrate
in a simple scheme how the full  
Wess-Zumino-Witten (WZW) functional \cite{Wess,Witten}
of a gauged chiral lagrangian in $D=4$ 
arises out of a pure gauge theory
of quark flavor in compactified $D=5$.
This model,  based upon
a $U(1)_L\times U(1)_R$ flavor
symmetry, discussed in \cite{HHH1}, mimics the chiral
structure of QCD, and was used to clarify how the counterterm
structure of the WZW functional arises in a parity-asymmetric gauging,
such as in the Standard Model.  

Here, however, we use it
to illustrate how the generic features of the holographic
origin of the chiral WZW term arise, where the 
$D=4$ mesons emerge out of the Wilson line
over the bulk gauge field, $A_5$.  
The general gauged WZW functional structure for $U(N)\times U(N)$
has been studied previously using deconstruction
\cite{cth1}, while the general gauged Kaymakcalan, Rajeev, and 
Schechter, (KRS) action \cite{KRS} has been derived from continuum 
compactification
of a pure $SU(N)$ Yang-Mills theory in detail in \cite{cth2}.
The present work is intended, in part, to clarify
the approach and results of \cite{cth2}.

We construct a manifestly $U(1)$ gauge-invariant theory in $D=5$.
The gauge fields propagate in the bulk,
with chiral quarks attached to chiral boundaries (branes), 
with $L$ ($R$) located at $x^5=0$ ($x^5=R$), respectively.
The quarks are chirally delocalized in $D=5$ \cite{cth3}, and their 
chiral anomalies \cite{anom} are nonzero on their respective boundaries, 
but would otherwise cancel if the boundaries were merged. 
 
The  boundary conditions on the $D=5$ gauge fields are subject to a 
minimal set of constraints: (I) there exists a massless physical $A_5$ 
zero mode, (or, more properly, a nontrivial Wilson line spanning the bulk)
which can be identified with chiral mesons;  and, (II)
there exists a tower of KK modes of the gauge fields, 
which is sufficiently rich, such that 
independently valued combinations of these fields 
exist on the boundary branes.

Much of what we cover in our pedagogical model is expected to apply to any 
theory of new physics in extra dimensions which satisfies (I-II) 
with chiral delocalization, including AdS $D=5$ models.
Irrespective of the specific $D=5$ geometry, our low-energy effective
theory  results are holographic, \ie, they are determined
at the boundary, as the integrands in 
the bulk involving the lower KK-modes are mostly {\em exact} expressions.
Since the theory we consider is arranged to be anomaly free,
the resulting effective action contains both the holographic
WZW, and a dual effective interaction in the bulk. 
This latter bulk interaction  takes
the form of a Chern-Simons (CS) functional \cite{Deseretal,Zumino:1983rz}
in the low energy effective theory variables \cite{cth2,cth3},
and cancels the anomalies on the boundaries.
 
With the chiral quarks attached to the boundaries, 
$\psi_L$ at $x^5=0$, and $\psi_R$ at $x^5=R$, respectively,
a ``constituent quark mass term'' 
is introduced of the form $m\bar{\psi}_L W\psi_R + h.c.$.
Here, $W$ is the Wilson line that spans the gap between the
boundary branes, and represents the dynamical chiral condensate
of the theory.  The Wilson line is identified with
a chiral field of mesons:
\beq
W(x^\mu) = \exp\!\left( \! i \! \! \int_0^{R} \! \! \! \! \! \! dx^5 
A_5(x^\mu ,x^5)\right) \equiv \exp\!\left( \frac{ia(x^\mu)}{f} \right ) .
\eeq
This is the rationale for requiring an $A_5$ zero mode, since we need 
that the pseudoscalar chiral meson $a$ be 
physical, and not be eaten by a KK-mode. 
The chiral symmetry breaking 
scale is specified by $f$, the ``pion decay constant";  
in any imitation of QCD chiral dynamics by an extra dimension,
chiral symmetry breaking is generically related to 
the compactification scale,
\ie, $f \sim 1/R$.

The quark anomalies \cite{anom}
on the boundary branes  must
be canceled by the anomalies that arise on the boundaries from
a bulk-filling Chern-Simons functional.
The anomaly cancellation condition determines the coefficient of
the Chern-Simons functional  \cite{Zumino:1983rz}.
We integrate out the quarks, taking the limit of large $m$.  In a special 
gauge, an effective ``Fujikawa'' action arises out of the quark 
Dirac determinant \cite{Fujikawa:1980eg}, which only involves the gauge 
fields on the boundary, and amounts to minus the Bardeen  counterterm
\cite{HHH1,KRS,anom}.

When the Chern-Simons term and the boundary term are added together, 
they yield a total effective action, $S^*$, all of whose anomalies cancel. 
Through a suitable reverse gauge transformation, we finally arrive at 
the purely bosonic anomaly-free action, 
consisting of a surface term and bulk term,  
\beq
S^* = \Gamma_{WZW}(a(x^\mu), A_L(x^\mu ), A_R(x^\mu )) + 
S_{CS}(A_A(x^\mu ,x^5 )). \label{tilS}
\eeq
$\Gamma_{WZW}(a,A_L, A_R)$
is the full (gauged) WZW functional, on the boundary. 
$S_{CS}$ is the original bulk Chern-Simons functional, whose anomalies 
cancel those of $\Gamma_{WZW}$.

For a purely $D=4$ theory such as QCD, we would have to discard the 
bulk term $S_{CS}$ and leave behind the WZW term $\Gamma_{WZW}$,
whose anomalies would then need be canceled by, e.g., suitable leptons, 
instead.
In a generic AdS-CFT correspondence setting, we would have both the
$\Gamma_{WZW} $ and the $S_{CS}$ effective
interactions to generate anomalies.  
$S_{CS}$, living in the bulk, also contains the KK modes
of the theory, and leads to a richer system of new degrees of freedom. 
These were previously studied in detail for QED in refs \cite{cth3,cth2},
whose basic procedure is effectively exemplified here.

We stress that this approach is significantly different from
the standard dimensional descent approach, pioneered
by Witten \cite{Witten}. In that approach,
{\em a chiral theory of  mesons 
in $D=5$} is considered from the start, and 
a $D=5$ chirally invariant closed but not exact  
pionic interaction,  $\Tr(d\pi d\pi d\pi d\pi d\pi)+...$ with a quantized 
coefficient is introduced. Dimensional descent to $D=4$ yields
the boundary term $\Tr(\pi d\pi d\pi d\pi d\pi)+...$, which is
subsequently gauged, {\em a posteriori}. 

By contrast, our present procedure 
begins with a pure gauge theory and
is {\em a priori} gauge invariant. The mesons emerge into being 
in the descent by compactification out of bulk gauge-field Wilson lines,
\cite{cth1}. We rely on the CS interaction to produce the full WZW term,
with its meson and gauge field components suitably linked---and possibly 
connected by further, still unappreciated, symmetries.
In this sense, our approach would fit within the broad dimensional descent 
framework of ref \cite{Zumino:1983rz}, conforming to its abstract mechanisms,
and has analogies with the bulk - string anomaly-cancellation 
mechanisms of \cite{Callan:1984sa}
and \cite{rest}, which rely on chiral zero modes of fermions on 
axionic-defect submanifolds. Our straightforward model is particularly 
revealing,  as now both mesons and gauge 
fields are part of the same construct.

We start by reviewing the pedagogical toy model and its canceling anomaly structure in 
$D=4$ in Section II. We proceed to derive the full WZW term exploiting 
interplay with the $D=5$ bulk in Section III, and discuss implications of the 
picture introduced in Section IV.

\section{Schematic Standard Model with chiral  pseudoscalar Interactions}
 
We imitate the standard model with a
simple $U(1)\times U(1)$ gauge theory, discussed
previously in ref \cite{HHH1}.
Consider a single color ($N_c=1$) and flavor  of ``quark,''
$q$, and a single ``lepton,'' $\ell$. Introduce $U(1)_L$
and $U(1)_R$ fundamental
gauge fields $A_L$ and $A_R$ into the quark action:
\beq
\label{q}
S_q = \int d^4x\; \bar{q}_L(i\slash{\partial} - \slash{A}_L)q_L 
+ \bar{q}_R(i\slash{\partial} - \slash{A}_R)q_R  . 
\eeq
Likewise, we gauge the ``lepton'' sector:
\beq
\label{ell}
S_\ell = \int d^4x\;
\bar{\ell}_L(i\slash{\partial} + \slash{A}_L)\ell_L 
+ \bar{\ell}_R(i\slash{\partial} + \slash{A}_R)\ell_R  . 
\eeq
Note the sign changes of the couplings. Taken together,
the gauge anomalies cancel between the quark and lepton sectors
in the $B-L$ currents,
\bea
\partial_\mu(\bar{q}\gamma^\mu q_L - \bar{\ell}\gamma^\mu \ell_L) = \partial^\mu
J^{B-L}_{L\mu} = 0  ,
\nonumber \\
\partial_\mu(\bar{q}\gamma^\mu q_R - \bar{\ell}\gamma^\mu \ell_R) = 
\partial^\mu J^{B-L}_{R \mu} = 0  .  
\eea
Anomalies remain in the ungauged $B+L$ currents, imitating the
structure of the standard model.
(The leptons are here only to cancel anomalies, and this role will be assumed, 
in the next section, by other, CS, bulk gauge interactions.)

One may see the analogy to hadronic physics and the chiral lagrangian of QCD. This may be done 
by decoupling the quarks made heavy by spontaneously
breaking the $U(1)_L\times U(1)_R$
symmetry to $U(1)$.  Doing so 
introduces an effective constituent quark mass term containing a pseudoscalar 
chiral Nambu-Goldstone boson,
\beq
m \bar{q}_L q_R e^{ia/f} + h.c.   ~.
\eeq
$a(x^\mu) /f$ is the analog in our system of $\pi^0/f_\pi$ in QCD. Under the 
$U(1)_L\times U(1)_R$ gauge transformations,
we have for the quark sector:
\bea
\label{gtq}
q_L & \rightarrow & e^{i\epsilon_L} q_L  \qquad
\delta A_L  = -d\epsilon_L   \qquad    \delta a = f\epsilon_L
\nonumber \\
q_R & \rightarrow & e^{i\epsilon_R} q_R  \qquad
\delta A_R  =  -d\epsilon_R   \qquad    \delta a = -f\epsilon_R.
\eea
(The leptons likewise transform as $\ell_L\rightarrow e^{-i\epsilon_L}\ell_L$
and $\ell_R\rightarrow e^{-i\epsilon_R}\ell_R$).
These gauge transformations are anomalous, shifting the original quark
sector action by the consistent anomalies \cite{Bardeen:1984pm},
\bea
\label{anom}
\delta S_q = \frac{1}{24\pi^2} \int d^4x \;( -\epsilon_L dA_L dA_L 
+ \epsilon_R dA_R dA_R). 
\eea
They are, of course, canceled by $\delta S_\ell $. We'll utilize 
compact differential form notation, \eg,
\beq
\int ABdC = \int d^4x\; \epsilon^{\mu\nu\rho\sigma}A_\mu B_{\nu} \partial_\rho C_{\sigma}.
\eeq

In principle, the large $m$ limit of integrating the quarks
out (which, for anomalies, in a sense, parallels confinement) 
results in an effective action \cite{dhokerI}, which is a functional
of the $U =  e^{ia(x^\mu) /f} $ and the gauge fields, $A_L$ and $A_R$,
$\Gamma_{WZW} (U, A_L, A_R)$.  This functional generates
the same anomalies as in eq.(\ref{anom}), and it is the WZW
functional  of our theory, which codifies the anomaly effects of the 
entire quark 
sector. 

However, in practice, it is easy to construct the parity-invariant WZW term 
heuristically,   by arranging a set of $D=4$ operators that generate
the independent $L$ and $R$ consistent anomalies.
One readily obtains 
\bea
\label{WZW1}
\Gamma_{WZW}  & = & -\frac{1}{24\pi^2} \int d^4 x 
\big[ A_R A_L dA_L - A_L A_R dA_R 
\nonumber \\
&& \!\!\!\! \!\!\!\! 
+  \frac{a}{f}(dA_LdA_L + dA_R dA_R + dA_L dA_R )
 \big]~.
\eea
Under the gauge transformations of eq.(\ref{gtq}), we
obtain precisely $\delta \Gamma_{WZW} = \delta S_q$ of eq.(\ref{anom}).
That is, $\Gamma_{WZW} $ is indeed the WZW part of the full action
of the low-energy hadronic theory   \footnote{Note that one could 
also formally obtain eq.(\ref{WZW1})  from the 
full $U(N)_L\times U(N)_R$ WZW functional specified by Kaymakcalan,
Rajeev and Schechter (KRS) \cite{KRS},
by taking its $U(1)_L\times U(1)_R$ limit. 
That full functional results out of coupling 
$\Gamma(U, 0,0)_{WZW} $ to gauge fields, but for the abelian subgroups in 
our limit, $\Gamma(U, 0,0)_{WZW} =0$.}. 
In the $f \rightarrow \infty$ limit in 
which the pseudoscalar $a(x^\mu) $ decouples, 
$\Gamma_{WZW} (\openone, A_L, A_R)\equiv \Gamma_0$ reduces to the 
Bardeen counterterm. (The role of the meson is no more visible in 
$\Gamma_{WZW} $ for $\epsilon_L=\epsilon_R $.)

In conclusion, the action $S_\ell+\Gamma_{WZW}$ is gauge invariant.

\section{Holographic origins of the WZW Term}

\subsection{The Setup}
Here, we derive eq.(\ref{WZW1}) by embedding the theory into
a $D=5$ $U(1)$ gauge theory. Denote $x^5\equiv y$. 
For simplicity, the bulk dimensionful constants have been absorbed 
into the definition of $y$ and $A_A$.
In this theory, 
we embed chirally delocalized
fermions on boundary  $D=4$-branes $I$ and $II$.
On brane I, at $y=0$, we have:
\beq
\label{qq}
S_I = \int_I d^4x\; \bar{q}_L(i\slash{\partial} - \slash{A}_L)q_L  , 
\eeq
and, on brane II at $y=R$, 
\beq
\label{qqqq}
S_{II} = \int_{II} d^4x\;  
\bar{q}_R(i\slash{\partial} - \slash{A}_R)q_R . 
\eeq
The $U(1)$ gauge field $A_A(x_\mu,y)$ propagates in the $D=5 $ bulk,
and
\beq
A_{\mu L} = A_\mu(x_\mu,0), \qquad \qquad A_{\mu R} = A_\mu(x_\mu,R).
\eeq

To cancel the anomalies in this 
theory, instead of $S_\ell$ above, we introduce into
the bulk the $D=5$ Chern-Simons functional  \cite{Zumino:1983rz}, 
\beq
S_{CS} = c\int d^4x \int_0^R dy \; \epsilon_{ABCDE}A^A\partial^B A^C 
\partial^D A^E ,
\eeq
where
$c$ is quantized by general arguments, \cite{Deseretal,Zumino:1983rz},  
\beq
c=\frac{1}{24\pi^2}  .
\eeq 
Further consider the abelian  Wilson line,
\beq
W = \exp \left(   i\int_0^R dy A_5(x_\mu , y )    \right) .
\eeq
This permits construction of  a bilocally gauge-invariant
fermion mass term,
\beq
S_m = -\int d^4x \; m \bar{q}_L(x_\mu) W q_R(x_\mu)+ h.c.
\eeq
We identify the line integral with the boundary meson,
\beq
\int_0^R dy A_5(x_\mu , y ) \equiv \phi(x_\mu)\equiv a(x_\mu) /f  ,
\eeq
even though, for formal convenience, below, we will utilize 
a bulk generalization, 
\beq
\int_0^y dz A_5(x_\mu , z ) \equiv \Phi(x_\mu,y),
\eeq
such that $\Phi(x_\mu,0)=0$, and $\Phi(x_\mu,R)=\phi(x_\mu)$ on the chiral 
boundaries. In the localized limit, $R 
\rightarrow 0$, $f\rightarrow \infty$, and the chiral boson drops out of the 
theory.  
\subsection{Anomalies} 

The actions $S_I$ and $S_{II}$ have anomalies on their respective branes.
Under the chiral boundary gauge transformations,
\bea
q_L\rightarrow e^{i\epsilon_L} q_L \qquad A_L \rightarrow A_L -d\epsilon_L
\nonumber \\
q_R\rightarrow e^{i\epsilon_R} q_R  \qquad A_R \rightarrow A_R -d\epsilon_R ,
\eea
we have:
\bea
\delta S_I = -\frac{1}{24\pi^2}  \int d^4x\; \epsilon_L  dA_L dA_L
\nonumber \\
\delta S_{II} = \frac{1}{24\pi^2}  \int d^4x\; \epsilon_R dA_R dA_R .
\eea
The mass term, as reviewed below, does not affect the 
full quark answer, 
\beq
\delta S_q= \delta S_I+ \delta S_{II}.
\eeq

In the bulk, the full gauge transformation is
\bea
\label{full}
A_A & \rightarrow & A_A - \partial_A \epsilon(x_\mu,y)
\nonumber \\
& &  \epsilon(x_\mu,0)=\epsilon_L ,
\qquad \epsilon(x_\mu,R)=\epsilon_R  .
\eea

It is useful to decompose the Chern-Simons term into
$A_5$ and $\partial_5=\partial_y$ components, expressed as 4-forms,
\beq
\label{decomposed}
S_{CS} = c\int d^4x dy \; [ 3A_5 dAdA - 2 A (\partial_y A) dA ]~.
\eeq
Under the gauge transformation
of eq.(\ref{full}),
$S_{CS}$ generates the anomalies on
the boundaries \cite{Deseretal,Zumino:1983rz}. (Naturally, there are
no anomalies in the bulk, as odd-dimensional fermions lack the chirality 
to produce such.) In what follows, 
it is useful to display this result
for the form chosen in eq.(\ref{decomposed}),
\bea
\label{GT}
\delta S_{CS} & = & 
c\int d^4x dy \;  \big[3(-\partial_y\epsilon) dAdA 
+ 2 (d\epsilon) (\partial_y A) dA
\nonumber \\
& & \qquad \qquad
+ 2 A (\partial_y d\epsilon) dA \big]
\nonumber \\
& = & 
-c\int d^4x dy \; \partial_y (\epsilon dAdA) 
\nonumber \\
& = &  \frac{1}{24\pi^2}\int d^4x \big[{\epsilon_L} dA_L dA_L
- {\epsilon_R} dA_R dA_R \big]\nonumber \\
& = &  -\delta S_q .
\eea
Thus, the full action, 
\beq
S^* = S_I + S_{II} + S_m + S_{CS} ,
\eeq
amounts to a gauge-invariant theory,
\beq
\delta S^* = 0  ,
\eeq
under the general gauge transformation,  eq.(\ref{full}).

Now note that the mesonless, localized ($R=0 \Leftrightarrow$ $f=\infty$) 
quark theory with $W=1$ is a $D=4$ action,
\bea
\label{dirac}
S_0 & = & \int d^4x\; \big[ \bar{q}_L(i\slash{\partial} - \slash{A}_L)q_L 
 + \bar{q}_R(i\slash{\partial} - \slash{A}_R)q_R
 \nonumber \\
 & & \qquad \qquad  - m(\bar{q}_Lq_R + h.c.)\big].
\eea
If we integrate out the fermions and expand in $1/m$
the logarithm of the resulting fermion determinant, 
$\Tr \log (\openone -i~\mathord{\not\mathrel{D}} / m)$, to leading 
order \cite{Fujikawa:1980eg}, 
we obtain an operator functional of $A_L$ and $A_R$
of the form,
\bea
\label{res}
\Gamma_0 & = &  -\frac{1}{6\pi^2} \int d^4x\; A V dV
\nonumber \\
& = &  \frac{1}{24\pi^2} \int d^4x\; \big[ A_L A_R dA_L + A_L A_R dA_R \big],
\eea
where $ A_R = V+A$ and 
$A_L = V-A$.

Thus, the sole surviving influence of the fermions removed from play is the
effective term, eq.(\ref{res}), which amounts to minus the Bardeen 
counterterm.  
If we added the Bardeen counterterm to $S_0$, we would cancel
$\Gamma_0$ to thus obtain a vanishing result.
This reflects the {\em anomaly decoupling}. 
The {\em covariant axial and vector currents} are defined by (I) adding
the Bardeen counterterm to the action and (II) then
varying the action with respect to 
either $V$ or $A$.
The  matrix elements of these currents can
be seen to 
vanish in the large $m$ limit. Therefore,
by adding the Bardeen counterterm to the action, and 
then integrating out
the massive fermions, the resulting action 
vanishes, and the current
matrix elements are not generated at all.

To be more precise, consider the covariant
axial anomaly equation: 
\beq
\partial_\mu j^{\mu 5} +2im\bar{q}\gamma^5q 
= \frac{1}{16\pi^2}(dVdV + \frac{1}{3}dAdA)  .
 \eeq
The matrix elements of this equation
are saturated on
the {\em lhs} by the $2im\bar{q}\gamma^5q $ term,
which survives in the
large $m$ limit; while
the matrix elements of $j^{\mu 5}$ are vanishing
like $1/m^2$.
These latter matrix elements, which would
be obtained by
variation with respect to $A$, are simply not generated
when $\Gamma_0$ is canceled by the counterterm.
 
\subsection{A Gauge Transformation} 
 
Our strategy is to exploit the gauge invariance 
of $S^*$,  and to go to an axial gauge, in which $A_5=0$ and $W=1$,
in which we can integrate the fermions out as above, and to then 
revert to the original gauge.

To this end, we introduce a gauge transformation in the bulk,
\beq
U(x, y)  = \exp \left( -i\int_0^y dz A_5(x_\mu , z )\right),
\eeq
\bea
B_A(x,y)& \equiv & A_A(x, y) - \partial_A \Phi(x,y)
\nonumber \\
\partial_A \Phi & = &  iU(x,y)^\dagger \partial_A U(x, y) ~,
\eea
and, on the boundaries, $\Phi(x,0) = 0$,
 $\Phi(x,R) = \phi(x)$ and 
\beq
\psi_L \rightarrow  U(x,0)\psi_L, 
\qquad
\psi_R \rightarrow  U(x,R)\psi_R .
\eeq

This transformation implies $B_5=0$, an axial gauge,
thus fixing $W=1$ in $S_m$, while $S_{CS}$ reduces to 
\beq
S^0_{CS} = -2c\int d^4x dy \;  B \partial_y B dB. \label{NCS}
\eeq
Thus,  we have
\beq
B_\mu(x,0)= A_{L\mu}(x), \quad B_{R\mu}(x) =A_{R\mu}(x) -\partial_\mu  
\phi(x). \label{voodoo}
\eeq
The gauge transformation chosen is parity-asymmetric, with $\phi$
appearing in $A_R$, but absent in $A_L$.
(We could have chosen a more parity-symmetric form,
\eg, defining $U=1$ at $y=1/2$,
but our present choice has the advantage of providing
an internal consistency check on
the interplay between the fermion loop ($\Gamma_0$) and the Chern-Simons
term in the final result.) The overall result will be parity symmetric.

On the $B_\mu=A_\mu- \partial_\mu \Phi(x,y) $ bulk components,  this gauge 
transformation has provided longitudinal components for
all of the massive $KK$-modes (unitary gauge), and allows
them to be treated as Stueckelberg fields, 
$B^n_\mu =A^n_\mu -\partial_\mu \phi^n$. However, significantly,  
the $y$-independent (zero) mode, which is the
Wilson line  integral over $A_5$, provides a residual 
uneaten $\phi$-field zero mode. 

\subsection{Derivation of the WZW Term}

With the effective removal of the phase in the Wilson line, $W$, we now
have the effective quark action,
\bea
S_0 & = & \int d^4x\; \bar{q}_L(i\slash{\partial} - \slash{B}_L)q_L 
 + \bar{q}_R(i\slash{\partial} - \slash{B}_R)q_R  ~.
 \nonumber \\
& &  - m(\bar{q}_Rq_L + h.c.).
\eea
Integrating out the quarks in the large $m$ limit, as reviewed above, we 
obtain the effective action from eq.(\ref{res}),
\beq
\label{res2}
\Gamma_0 
 =  \frac{1}{24\pi^2} \int d^4x\; \big[ B_L B_RdB_L + B_L B_RdB_R \big] .
\eeq
This is, of course, not separately gauge invariant.
The gauge invariant action is 
\beq
S^*= \Gamma_0 +  S^0_{CS}. 
\eeq

Now, performing the 
{\em inverse} gauge transformation, (\ref{voodoo}), from $B$ to $A$, yields, 
on the one hand, 
\bea
\label{one}
\Gamma_0  & = & 
 \frac{1}{24\pi^2} \int d^4x\; \big[ A_L A_RdA_L + A_L A_R dA_R
 \nonumber \\
& &+d\phi  ( A_L dA_L + A_L dA_R)\big ].
\eea
Note the appearance of $d\phi$ with $A_R$, but not $A_L$,
a consequence of our parity-asymmetric gauge transformation choice,
so the expression is not parity symmetric.

On the other hand, the gauge-variant bulk term (\ref{NCS}) 
now also transforms under the same inverse transformation to  
\beq
S^0_{CS} = -2c\int d^4x dy \;  (A-d\Phi) \partial_y (A-d\Phi) dA. 
\eeq
This expression may be recast through $\partial_y \Phi= A_5$ as
\beq
S^0_{CS}\! =\! -c\!\!\int\!\!\! d^4x dy [ 2A (\partial_y A- d A_5) dA  
+\Phi \partial_y  (dA dA)] .  
\eeq
Through an integration by parts, we recover the bulk term (\ref{decomposed})
and an additional boundary term, 
 \bea
S^0_{CS}&=& c\int d^4x dy \; ( 3A_5 dAdA - 2 A (\partial_y A) dA )\nonumber \\
&-& c\int d^4x dy \; \partial_y ( \Phi dA dA) \nonumber \\
&=& S_{CS} + \delta S^0.
\eea
The boundary piece is also parity asymmetric, 
\beq
\delta S^0 = -\frac{1}{24\pi^2}\int d^4x \big[ {\phi} dA_R dA_R \big], 
\eeq
and, when added to above fermion determinant boundary contribution, (\ref{one}), 
yields the full, parity-symmetric, gauge-noninvariant  
WZW action, (\ref{WZW1}), on the boundary branes, 
\bea
\Gamma_{WZW} & = & -\frac{1}{24\pi^2} \int d^4 x \;
\big[ A_R A_L dA_L - A_L A_R dA_R \nonumber \\
 & &+    \phi (dA_LdA_L + dA_R dA_R + dA_L dA_R ) 
\big].
\eea
To restore proper dimensions, take $\phi=a/f$.

So it is the interplay of the fermion determinant with the holographic 
gauge-invariance violation of the bulk CS term which gives rise to the 
full gauge-invariant effective action. This action then, 
\beq
S^*=\Gamma_{WZW} + S_{CS}, 
\eeq
amounts to the 
boundary WZW term and the 
bulk CS term, which cancel each other's anomalies \footnote{Strictly speaking, 
in principle, the above straighforward derivation could 
have been duplicated by brute force application of the procedure of 
ref \cite{dhokerI}, \ie, direct integration out of the quarks, 
although the presence of the pseudoscalar meson would have complicated 
the evaluation of the functional determinant.}. 

Consequently, 
either the boundary or (minus) the bulk terms are equivalent bosonic 
representations of the very same boundary fermion anomalies, and may 
thus be effectively regarded as holographic duals to each other \cite{cth1}.

\section{Implications} 

The significance of dimensionally-descending bosonic functionals in anomaly 
physics has been long appreciated \cite{Witten,Zumino:1983rz}. Moreover,
even though superficially very different, the CS functionals in odd 
spacetime dimensions and the WZW functionals in even ones are, in fact,
linked in a unique structure \cite{cth1}. 

Nevertheless, within the context of 
actual brane models, the ready holographic cancellation of anomalies by the 
respective bulk CS and boundary WZW functionals identified here, and, indeed, 
the effective dual representation of the very same anomalies, is 
particularly transparent, and perhaps useful to model-building. 
Loosely, one may think of the WZW term as the holographic AdS/CFT image of 
the CS functional. 

To be sure, the present simple model deals with 
anomaly inflows specified by fermions strictly confined to boundary branes.
Given the holographic projection identified here, however, one is justified to 
ask whether fermions {\em incompletely localized} on boundary 
branes \cite{Callan:1984sa,rest},
developing chiral zero modes on the boundary,  
would {\em also} be
amenable to analogous treatment, if integrated out to a bosonic effective
action. 
In odd spacetime dimensions, there are no chiral fermions leading to anomalies,
but they may develop chiral zero modes on even-dimensional submanifolds.
In such a theory, would the higher KK modes, which did not couple to 
the WZW functional here (but which are present, inertly, in the CS 
functional in our present treatment), insinuate themselves into a 
generalization of the full (gauged) 
WZW term derived? The issue is under current investigation.
   
\vskip 0.2in
\noindent
{\bf Acknowledgments}
\vskip 0.1in
\noindent
We thank Jeff Harvey and Richard Hill 
for helpful discussions.
Research supported by the U.S.~Department of Energy  
grant DE-AC02-76CHO3000, and by the U.S.~Department of Energy, 
Division of High Energy Physics, Contract DE-AC02-06CH11357 as well as the 
University of Chicago -- Argonne Joint Theory Institute.

\end{document}